\documentclass[epj]{svjour}
\usepackage{graphicx}

\def\i{\mathrm{i}}
\def\e{\mathrm{e}}
\def\d{\mathrm{d}}
\def\half{{\textstyle{1\over2}}}
\def\thalf{{\textstyle{3\over2}}}
\def\h{{\scriptscriptstyle{1\over2}}}
\def\th{{\scriptscriptstyle{3\over2}}}

\def\vec#1{\mbox{\boldmath$#1$}}
\def\svec#1{\mbox{{\scriptsize \boldmath$#1$}}}

\def\CG#1#2#3#4#5#6{C^{#5#6}_{#1#2#3#4}}

\begin{document}

\title{Direct calculation of the $K$ matrix
for pion electro-production in the delta channel}

\author{%
P.~Alberto\inst{1} 
\and
L.~Amoreira\inst{2} 
\and
M.~Fiolhais\inst{1} 
\and
B.~Golli\inst{3}
\and
S.~\v{S}irca\inst{4}}

\institute{%
Department of Physics and Centre for Computational Physics,
                  University of Coimbra,
                  3004-516 Coimbra, Portugal
\and
Department of Physics,
                  University of Beira Interior,
                  6201-001 Covilh\~a, Portugal
\and
Faculty of Education,
              University of Ljubljana and J.~Stefan Institute,
              1000 Ljubljana, Slovenia
\and
Faculty of Mathematics and Physics and J.~Stefan Institute,
              University of Ljubljana,
              1000 Ljubljana, Slovenia}

\date{\today}

\abstract{%
We present a method to calculate directly the $K$~matrix
for the pion electro-production processes in the framework
of chiral quark models which allows for a clean separation
of the resonant amplitudes from the background.
The method is applied to the calculation of the multipole
amplitudes $M_{1+}$, $E_{1+}$, and $S_{1+}$ in the $\Delta$
channel within the Cloudy Bag Model.  A good overall
description is found in a broad energy range.}

\PACS{12.39.-x, 13.40.Gp, 13.60.Le}

\maketitle

\section{Introduction}

Electro-production experiments reveal important information 
on the structure of nucleon resonances and provide
stringent tests of quark models.  In particular the $\Delta(1232)$
has been studied extensively (see \cite{Tiator03} and references
therein for a short review and basic nomenclature).  
In these studies (\cite{FGS}, see also \cite{gellas,satolee}
and \cite{Walcher} in the elastic sector) the important role of 
the pion cloud in baryons has become evident, manifesting itself
in a relatively large probability for the quadrupole excitation 
of the $\Delta$.
Such a large probability cannot be explained in the framework of the
constituent quark model unless the exchange currents generated from
the one-pion-exchange and/or the one-gluon-exchange potentials are
included as required by current conservation \cite{Buchmann}.
This is also an indication of the relevance of pions or, 
equivalently, the $q\bar{q}$ pairs.

In most approaches only the amplitudes for the excitation of
the resonance have been calculated, treating it as a bound state,
{\it i.e.\/} ignoring its decay.  
While such an approach can be justified in the case of weak 
or electro-magnetic resonance decays, its use in the case 
of strongly decaying resonances is not well founded.
In fact, the excited states manifest themselves as resonances
in meson scattering and, since the resonant scattering (as 
well as the electro-production process) is always accompanied
by non-resonant processes, the extraction of the resonant
amplitudes is not straightforward.  The resonant contribution
is related to the pole residue of the corresponding $K$~matrix; 
following the notation of \cite{Davidson90}, the $K$~matrix
for scattering is parameterized as
\begin{equation}
   K = {C\over E_\mathrm{R} - E} + D
\label{Kscatter}
\end{equation} 
and the $K$~matrix for the electro-production 
as
\begin{equation}
   K = {A\over E_\mathrm{R} - E} + B\,,
\label{Kepro}
\end{equation} 
where $C$ and $A$ represent the resonant parts and $D$ and $B$ 
the background.  
The $E$ is the invariant mass of the system.  
In order to extract the resonant part of electro-production 
amplitudes of given multipolarities, the information not only from 
electro-production but also from scattering is needed.  
In the model calculation of these amplitudes one usually takes 
the experimental values for the parameters of the resonance 
such as the position, the width, and the background phase shift.  
While this is possible in the case of the $\Delta(1232)$ where 
relatively precise measurements are available, such an approach cannot
be used in the case of other resonances, {\it e.g.\/} the Roper.
The only sensible approach is therefore to calculate both
electro-production and scattering within the same model.

The aim of this work is to construct a feasible computational
scheme for the full electro-production amplitudes, calculating
directly the pertinent $K$~matrices.
The resulting matrices for scattering and electro-production
appear in the forms (\ref{Kscatter}) and (\ref{Kepro});
to separate the resonant contribution from the background 
it is therefore sufficient to pick up the respective residues.
From the $K$~matrices it is possible to deduce the electro-production 
amplitudes as a function of $E$, as well as their dependence
on the photon virtuality $Q^2$.
Furthermore, the method is able to predict the EMR and CMR ratios
not only at the $K$~matrix pole but also at the $T$~matrix pole
which allows us to make the comparison with calculations based
on the $T$~matrix.

We show that in models in which the pion field is linearly coupled 
to the quark core it is possible to construct a computational scheme 
which goes beyond the usual perturbation approach.
We present the calculation for one such model, the Cloudy Bag Model
in sec.~\ref{CBM}.  
The amplitudes are sufficiently well reproduced from the pion 
threshold up to the energy region where the two-pion decay becomes 
important and the assumption of the single-pion channel breaks down. 
Our calculation of the $M_{1+}$ amplitude is similar to
that in ref.~\cite{Weyrauch} using the $T$-matrix approach
(see also \cite{kalber}).  However, to the best of our knowledge neither
the full $E_{1+}$ amplitude has been calculated in the framework
of quark models, nor has the $Q^2$-dependence of the amplitudes
been explored away from the resonance.

\section{Electro-production amplitudes in the $K$-matrix formalism}

The $K$~matrix for $\pi$N scattering is defined as:
$$
K_{\beta\alpha} = -\pi \langle \Phi_\beta|H'|\Psi_\alpha^\mathrm{P}\rangle
                = -\pi \langle \Psi_\beta^\mathrm{P}|H'|\Phi_\alpha\rangle
$$
(see {\it e.g.\/} \cite{Newton}),
where $H'$ is the interaction part of the Hamiltonian, 
$|\Phi_\alpha\rangle$ are the asymptotic (unperturbed) states 
with $\alpha$ labeling the pion-nucleon system, and
$|\Psi_\alpha^\mathrm{P}\rangle$ are the {\em principal-value\/} states
satisfying
\begin{equation}
|\Psi_\alpha^\mathrm{P}\rangle = |\Phi_\alpha\rangle
  + {\mathcal{P}\over E-H_0}\,H'|\Psi_\alpha^\mathrm{P}\rangle \,,
\label{Pstate}
\end{equation}
and normalized as
$$
  \langle\Psi_\alpha^\mathrm{P}(E)|\Psi_\beta^\mathrm{P}(E')\rangle = 
  \delta(E-E')\delta_{\alpha\beta}(1 + K^2)_{\alpha\alpha}\,.
$$
The $K$~matrix is related to the familiar $T$~matrix\footnote{%
Here we use the definition of the $T$~matrix as {\it e.g.\/} in 
\cite{Davidson90} which differs by a factor $\pi$ from that in 
\cite{Newton}.} 
by
$$
   T = -{K\over 1 -\i K}\,.
$$
In the case of a single channel, the $K$~matrix is equal to the 
tangent of the $\pi\mathrm{N}$-scattering phase shift, $K=\tan\delta$.

In order to introduce the electro-production amplitudes in this formalism,
we make the usual assumption that ``switching on'' the electro-magnetic
interaction $H_\gamma$ does not change the strong scattering amplitudes,
{\it i.e.\/} the principal-value states (\ref{Pstate}) remain unchanged.
The $K$~matrix for the electro-magnetic process is
$$
   K_{\gamma\pi} = 
   -\pi  \langle \Psi^\mathrm{P}(m_s,m_t;\vec{k}_0,t)|H_\gamma
     |\mathrm{N}(m_s',m_t');\vec{k}_\gamma,\mu\rangle \>.
$$
Here the initial state corresponds to the incoming
virtual photon with four-momentum $(\omega_\gamma,\vec{k}_\gamma$), 
$\omega_\gamma^2-\vec{k}_\gamma^2 = -Q^2$ and polarization $\mu$,
and the nucleon with the third component of spin $m_s'$ and isospin
$m_t'$;
the final state consists of a nucleon and a scattered pion with 
four-momentum ($\omega_0,\vec{k}_0$) and third component of isospin $t$.
In the c.m. frame the nucleon momentum is opposite to the photon
(pion) momentum, $\vec{k}_\gamma$, which defines the direction of the
$z$-axis.

We expand the pion-nucleon states in a basis with good total 
angular momentum $J$ and isospin $T$ which we write as
$$
  \Psi^\mathrm{P}_{JT}(M_JM_T;k_0,l)
   = K_{\pi\pi}^{JT}\,\widetilde{\Psi}_{JT}(M_JM_T;k_0,l)\,.
$$
Here $K_{\pi\pi}^{JT}$ is the $K$~matrix for pion scattering in the 
channel $JT$ and is related to the corresponding $T$~matrix by
$T_{\pi\pi}^{JT} = K_{\pi\pi}^{JT}/(1-\i K_{\pi\pi}^{JT})$.
The advantage of using $\widetilde{\Psi}$ over $\Psi^\mathrm{P}$ is that 
it is a smooth function of the energy and its norm does not diverge 
at a (possible) resonance where $K\equiv\tan\delta\to\infty$.
The incoming photon-nucleon state takes the form
$$
  |\mathrm{N}(m_s',m_t');\vec{k}_\gamma,\mu\rangle
   = \sqrt{\omega_\gamma k_\gamma}\,a^\dagger_\mu(\vec{k}_\gamma)
                                    |\mathrm{N} m_s' m_t'\rangle\,,
$$
where $a^\dagger_\mu(\vec{k}_\gamma)$ is the creation operator 
for the photon and the factor $\sqrt{\omega_\gamma k_\gamma}$
ensures proper normalization.

In this article we study the production of $p$-wave pions 
in the $\Delta$ channel below the two-pion threshold, though
the calculation can actually be extended to higher energies 
until the effect of the two-pion channel becomes prominent.
For simplicity, we neglect the recoil corrections to the nucleon 
ground state.  
To obtain the  electro-production amplitudes in this channel,
we keep only the $p$-wave pions and the $J=T=\thalf$
components in the expansion of the $\pi$N system
(in this case we drop the $JT$ superscripts).
The $T$~matrix for electro-production can then be written as
$$
 T_{\gamma\pi} =
   \pi T_{\pi\pi} \,{1\over\sqrt{2\pi}^3}\,\sum_m
        K_\lambda \, Y_{1m}(\hat{r})\,
      \CG{\h}{m_s}{1}{m}{\th}{\lambda}\CG{\h}{\h}{1}{0}{\th}{\h}\,.
$$
Here we have introduced the analogues of the familiar transverse 
helicity amplitudes:
\begin{eqnarray}
    K_\lambda &=&
    \sqrt{\omega_\gamma k_\gamma}\,
    \langle \widetilde{\Psi}_\Delta(M_J=\lambda)|
    {e_0\over\sqrt{2\omega_\gamma}}
    \int\d\vec{r}\,\vec{\varepsilon}_\mu\cdot\vec{j}(\vec{r})
\nonumber \\ && \times
    \e^{\i\svec{k}_\gamma\cdot\svec{r}}
        |\mathrm{N}(m_s'=\lambda-\mu)\rangle  \,,
\label{Klambda}
\end{eqnarray}
where $\vec{j}(\vec{r})$ is the vector part of the electro-magnetic 
current and $M_T=m_t'=\half$.
The transverse electro-production amplitudes are
\begin{equation}
 M_{1+}^{(3/2)} 
    = -T_{\pi\pi}\sqrt{3 \over 16 k_0k_\gamma}\,
      {1\over2\sqrt{3}}\,(3K_{3/2} + \sqrt{3}K_{1/2}) \,,
\label{M1plus}
\end{equation}
\begin{equation}
 E_{1+}^{(3/2)} 
    = T_{\pi\pi}\sqrt{3 \over 16 k_0k_\gamma}\, 
      {1\over2\sqrt{3}}\,(K_{3/2} - \sqrt{3}K_{1/2}) \,.
\label{E1plus}
\end{equation}
The scalar amplitude is
$$
 S_{1+}^{(3/2)} 
      = T_{\pi\pi}\sqrt{3 \over 16 k_0k_\gamma}\,
        {1\over\sqrt{2}}\,K_S \,,
$$
where
\begin{equation}
    K_S = e_0\sqrt{k_\gamma\over2}
    \langle \widetilde{\Psi}_\Delta(M_J=\half)|
    \int\d\vec{r}\,
    \rho(\vec{r})
    \e^{\i\svec{k}_\gamma\cdot\svec{r}}
        |\mathrm{N}(m_s'=\half)\rangle\,.
\label{KL}
\end{equation}
The longitudinal amplitude $L_{1+}$ is obtained by simply 
replacing the density operator by 
$\vec{\varepsilon}_0\cdot\vec{j}(\vec{r})$.

The differential cross-section averaged over the initial 
states $m_s'=\pm\half$ and $\mu=\pm1$ reads
\begin{equation}
  {\d\sigma_T\over\d\Omega} = {(2\pi)^4\over k_\gamma^2}\,
{1\over4}\sum_{m_s'\mu}\left|{T_{\gamma\pi}\over\pi}\right|^2
\label{sigmaT}
\end{equation}
for the transverse photons; for the longitudinal photons the
average is taken only over one polarization, $\mu=0$.
Equation~(\ref{sigmaT}) yields the familiar expression in terms of 
the pertinent electro-production amplitudes and the scattering angle
(see {\it e.g.\/} \cite{DT}).
The EMR and CMR ratios are defined in the usual way \cite{pospischil}
as
\begin{eqnarray}
\mathrm{EMR} &=& {\mathrm{Re}\,[\,E_{1+}^{(3/2)\ast} M_{1+}^{(3/2)}\,]
           \over |\,M_{1+}^{(3/2)}\,|^2} \,,
\nonumber\\
\mathrm{CMR} &=& {\mathrm{Re}\,[\,S_{1+}^{(3/2)\ast} M_{1+}^{(3/2)}\,]
           \over |\,M_{1+}^{(3/2)}\,|^2} \,.
\nonumber
\end{eqnarray}

\section{Calculation of the $K$~matrix in chiral quark models}

In this work we consider quark models in which $p$-wave pions 
couple linearly to the three-quark core.
Assuming a pseudo-scalar quark-pion interaction, the part of 
the Hamiltonian referring to pions can be written as
\begin{eqnarray}
 H_\pi &=& 
  \int\d k \sum_{mt}\left\{\omega_k\,a^\dagger_{mt}(k)a_{mt}(k)
 \right. 
\nonumber\\ && \left.
    + \left[V_{mt}(k) a_{mt}(k) 
        + V_{mt}^\dagger(k)\,a^\dagger_{mt}(k)\right] \right\},
\label{Hpi}
\end{eqnarray}
where
$a^\dagger_{mt}(k)$ is the creation operator for a $p$-wave
pion with the third components of spin $m$ and isospin $t$, and
\begin{equation}
      V_{mt}(k) = -v(k)\sum_{i=1}^3 \sigma_m^i\tau_t^i
\label{Vmt}
\end{equation}
is the general form of the pion source, with $v(k)$ depending on 
the particular model.

Chew and Low \cite{ChewLow} considered a similar model as (\ref{Hpi}) 
except that they did not allow for excitations of the nucleon core.
They showed that the $T$~matrix for $\pi$N scattering is proportional 
to $\langle\Psi^{(-)}(E)|V_{mt}(k)|\Phi_\mathrm{N}\rangle$,
where $\Psi^{(-)}(E)$ are the incoming states.
In general, the corresponding formula for the $K$~matrix cannot be
written in such a simple form.
However, in the $JT$ basis, in which the $K$ and $T$~matrices are 
diagonal, it is possible to express the $K$~matrix in the form\footnote{%
In the static approximation, $k_0$ is uniquely related to the energy 
$E=E_\mathrm{N}+\omega_0$, so one can use either $k_0$ or $E$ 
to label the states; for the on-shell $K$~matrix we write
$K(k_0,k_0) = K(E)$.}
\begin{equation}
   K_{\pi\pi}^{JT}(k,k_0) =  -\pi\sqrt{\omega_k\over k}
             \langle\Psi^\mathrm{P}_{JT}(E)||V(k)||\Phi_\mathrm{N}\rangle\,.
\label{KCL}
\end{equation}
The corresponding principal-value state obeys a similar 
equation as the in- and out-going states in the Chew-Low model:
\begin{equation}
  |\Psi^\mathrm{P}_{JT}\rangle = \sqrt{\omega_0\over k_0}\left\{
   \left[a^\dagger(k_0)|\Phi_\mathrm{N}\rangle\right]^{JT}
  - {\mathcal{P}\over H-E}\,
  \left[V(k_0)|\Phi_\mathrm{N}\rangle\right]^{JT}
    \right\},
\label{EoMCL}
\end{equation}
where $[\kern3pt]^{JT}$ denotes coupling to good $J$ and $T$.  
In order to rewrite this equation in a form more suitable
for a practical calculation, 
we insert into (\ref{EoMCL}) the complete set of eigenstates of $H$
\begin{eqnarray}
  \mathbf{1} &=& |\Phi_\mathrm{N}\rangle\langle\Phi_\mathrm{N}|
           +\sum_{JT}\int_{E_\mathrm{N}+m_\pi}^\infty \kern-6pt\d E\,\, 
               {|\Psi^\mathrm{P}_{JT}(E)\rangle\langle\Psi^\mathrm{P}_{JT}(E)|
                               \over 1 + {K_{\pi\pi}^{JT}}(E)^2} 
\nonumber\\ &&
+ \,2\pi\hbox{-states}\, 
          + \cdots
\nonumber
\end{eqnarray}
For energies below the 2-pion threshold only the one-pion
states contribute, hence the equation of motion takes the form:
\begin{eqnarray}
  |\Psi^\mathrm{P}_{JT}(E)\rangle &=& \sqrt{\omega_0\over k_0}
  \left[a^\dagger(k_0)|\Phi_\mathrm{N}\rangle\right]^{JT}
\nonumber\\ && \kern-48pt -
 \int\d E'\, {|\Psi^\mathrm{P}_{JT}(E')\rangle\over1 + K(E')^2}
    {\langle\Psi^\mathrm{P}_{JT}(E')|
            \left[V(k_0) |\Phi_\mathrm{N}\rangle\right]^{JT}\over E'-E}\,.
\label{EoM4Psi}
\end{eqnarray}

Let us remark that for a general chiral quark model, 
the $K$~matrix and the corresponding principal-value state
can be calculated variationally using the Kohn variational principle.
For the single-channel scattering of a meson with momentum
$k_0$ and energy $\omega_0$ \cite{varK} it amounts to requiring
the stationarity of
$$
  \tan\delta - {\pi\omega_0\over k_0}\,
         \langle \Psi^\mathrm{P}|H-E|\Psi^\mathrm{P}\rangle \,,
$$
where  $\Psi^\mathrm{P}$ is a suitable chosen trial state.

\section{Solution in the $\Delta$ channel}

The important difference between our approach and
the approach of Chew and Low is that the interaction $V(k)$ can 
generate bare quark states with quantum numbers different from the 
ground state by flipping the spin and isospin of the quarks.
Furthermore, in the same spirit one can consider a more general
type of models in which the quarks can be excited to higher
spatial states.
The state with the flipped spins plays a crucial role in the formation 
of the resonance in the delta channel.
The general form (\ref{EoM4Psi}) therefore suggests the following 
ansatz
in which we separate the resonant quasi-bound state $\Phi_\Delta$
from the state corresponding to pion scattering on the nucleon:
\begin{eqnarray}
  |\Psi_\Delta\rangle &=& \sqrt{\omega_0\over k_0}\left\{
\left[a^\dagger(k_0)|\Phi_\mathrm{N}\rangle\right]^{\th\th}\right.
\nonumber\\ && \left. \kern-36pt
  + \int\d k\,\,{\chi(k,k_0) \over \omega_k-\omega_0}
  \left[a^\dagger(k)|\Phi_\mathrm{N}^E(k)\rangle\right]^{\th\th}
  + c_\Delta^E|\Phi_\Delta\rangle\right\}.
\label{PsiD}
\end{eqnarray}
We require that the resonant state $\Phi_\Delta$ does not contain
components with pions around the nucleon, since such a component
is already included in the first two terms.
We therefore impose the following constraint on $\Phi_\Delta$:
$$
  \langle\Phi_\Delta|a^\dagger_{mt}(k)|\Phi_\mathrm{N}\rangle = 0\,. 
$$
We allow for the modification of the pion cloud in the nucleon in 
the presence of the scattering pion but require that such a state,
$|\Phi_\mathrm{N}^E\rangle$, asymptotically goes over to the 
true ground state $|\Phi_\mathrm{N}\rangle$.  
The pion amplitude is related to the $K$~matrix by
\begin{equation}
  \chi(k_0,k_0) = {k_0\over\pi\omega_0}\,K_{\pi\pi}(k_0,k_0)\,.
\label{K2chi}
\end{equation}

Iterating (\ref{EoM4Psi}) using the ansatz (\ref{PsiD})
we obtain the solution for $\chi(k,k_0)$ in the form
$$
\chi(k,k_0) = -c^E_\Delta\mathcal{V}_{\Delta \mathrm{N}}(k) 
+ \mathcal{D}(k_0,k)\,.
$$
The $\mathcal{V}_{\Delta \mathrm{N}}(k)$ and $\mathcal{D}(k_0,k)$
obey the integral equations: 
\begin{eqnarray}
\mathcal{V}_{\Delta \mathrm{N}}(k) &=&  {V}_{\Delta \mathrm{N}}(k)
  +\int{\d k'\over\omega_k'-\omega_0}\,\mathcal{K}_\mathrm{N}(k,k')
   \mathcal{V}_{\Delta \mathrm{N}}(k')\,, 
\nonumber \\
\mathcal{D}(k_0,k)  &=& \mathcal{K}_\mathrm{N}(k_0,k)
  +\int{\d k'\over\omega_k'-\omega_0}\,\mathcal{K}_\mathrm{N}(k,k')
  \mathcal{D}_\mathrm{N}(k_0,k') \,,
\nonumber
\end{eqnarray}
where
$$
 {V}_{\Delta \mathrm{N}}(k) =
 \langle\Phi_\Delta||V(k)||\Phi_\mathrm{N}\rangle \,,
$$
and $\mathcal{K}_\mathrm{N}(k_0,k)$ is the kernel involving 
scattering channels also for $JT\ne\thalf\thalf$.
It is dominated by the crossed term involving the nucleon;
the contributions from the crossed terms involving the delta
and the Roper resonance are small while the channels with
$J\ne T$ negligible.
Neglecting the widths of the resonances (see discussion in appendix~A)
as well as assuming $\Phi_\mathrm{N}^E\approx\Phi_\mathrm{N}$
allows us to write the kernel in the form
\begin{eqnarray}
\mathcal{K}_\mathrm{N}(k',k) &=&  {4\over9}\,
  {\langle\Phi_\mathrm{N}||V(k')||\Phi_\mathrm{N}\rangle 
   \langle\Phi_\mathrm{N}||V(k)||\Phi_\mathrm{N}\rangle
    \over \omega_k + \omega_k' - \omega_0}
\nonumber\\
&&\kern-9pt
+   {1\over36}\, {\langle\Phi_\mathrm{N}||V(k')||\Phi_\Delta\rangle
   \langle\Phi_\mathrm{N}||V(k)||\Phi_\Delta\rangle
    \over \omega_k + \omega_k'+ \varepsilon_\Delta -\omega_0}
\nonumber\\
&&\kern-9pt
+ {4\over9}\,
  {\langle\Phi_\mathrm{N}||V(k')||\Phi_\mathrm{R}\rangle 
   \langle\Phi_\mathrm{N}||V(k)||\Phi_\mathrm{R}\rangle
    \over \omega_k + \omega_k' +\varepsilon_\mathrm{R} - \omega_0}\,.
\label{KN}
\end{eqnarray}
Here $\varepsilon_\Delta= E_\Delta-E_\mathrm{N}$ 
and $\varepsilon_\mathrm{R}= E_\mathrm{R}-E_\mathrm{N}$ are the
delta-nucleon and the Roper-nucleon energy splittings, respectively.

The solution for $c^E_\Delta$ can be written as
$$
\left[E_\Delta(\omega_0) - E\right] c^E_\Delta 
      = -\mathcal{U}_{\Delta \mathrm{N}}(k_0)
$$
with
$$
E^\Delta_\Delta=\langle\Phi_\Delta|H|\Phi_\Delta\rangle
$$
and
\begin{eqnarray}
\mathcal{U}_{\Delta \mathrm{N}}(k_0) &=& {V}_{\Delta \mathrm{N}}(k_0)
 +\int{\d k\over\omega_k-\omega_0}\,{V}_{\Delta \mathrm{N}}(k)
       \mathcal{D}_\mathrm{N}(k_0,k)\,,
\nonumber\\
E_\Delta(\omega_0) &=& E_\Delta^\Delta - \Sigma_\Delta(\omega_0)
 = E_\Delta
  + \Sigma_\Delta(\varepsilon_\Delta) - \Sigma_\Delta(\omega_0)\,, 
\nonumber\\
\nonumber
\Sigma_\Delta(\omega_0) &=& 
  \int{\d k\over\omega_k-\omega_0}\,\mathcal{V}_{\Delta \mathrm{N}}(k)\,
         {V}_{\Delta \mathrm{N}}(k)\,,
\nonumber
\end{eqnarray}
where $E_\Delta=E_\Delta(\omega_0=\varepsilon_\Delta)$ is the position 
of the pole (of the $K$~matrix).
In a practical calculation we can always adjust a model parameter
({\it e.g.\/} the bare $\Delta$ energy) such that 
$E_\Delta$ corresponds to the experimental value.

The final result for the $K$~matrix, in which the resonant and the 
background contributions are explicitly separated, is
\begin{eqnarray}
 K_{\pi\pi}(E)  &=& \tan\delta = \pi\,{\omega_0\over k_0}\,\chi(k_0,k_0) 
\nonumber\\ 
&=&    \pi\,{\omega_0\over k_0}\left[
  {\mathcal{U}_{\Delta \mathrm{N}}(k_0)\mathcal{V}_{\Delta \mathrm{N}}(k_0) 
  \over E_\Delta(\omega_0) - E}
  + \mathcal{D}(k_0,k_0)\right].
\nonumber
\end{eqnarray}

Having obtained the parameters of the scattering state (\ref{PsiD}),
the calculation of the electro-production amplitudes is straightforward.
In the type of models we are considering here,
the current and the charge density operators can be split into 
quark and pion parts:
\begin{eqnarray}
  \vec{j}(\vec{r})  &=&
  \bar{\psi}\vec{\gamma}({\textstyle{1\over6}} + \half\tau_0)\psi
  + \i \sum_t t \pi_t(\vec{r})\vec{\nabla}\pi_{-t}(\vec{r})\; ,
\label{current}\\
  \rho(\vec{r}) & = &
  \bar{\psi}\gamma_0({\textstyle{1\over6}} + \half\tau_0)\psi
  - \i \sum_t t \pi_t(\vec{r}) P^\pi_{-t}(\vec{r})\;,
\label{charge}
\end{eqnarray}
where $P^\pi$ stands for the canonically conjugate pion field.
The procedure used to calculate the matrix elements of (\ref{Klambda})
and (\ref{KL}) is sketched in appendix~A.

\section{Results for the Cloudy Bag Model}
\label{CBM}

We shall investigate the capability of the method by calculating
the electro-pro\-duc\-tion amplitudes $M_{1+}$, $E_{1+}$, 
and $S_{1+}$ in the resonant $J=T=\thalf$ channel in 
the framework of the Cloudy Bag Model.
The Hamiltonian of the model has the form (\ref{Hpi}) and (\ref{Vmt}) 
with
$$
  v(k) = {1\over2{f_\pi}}\,{k^2\over\sqrt{12\pi^2\omega_k}}\,
    {\omega^0_\mathrm{MIT}\over\omega^0_\mathrm{MIT}-1}\,
            {j_1(k{R})\over  k{R}}\,,
$$
where $\omega^0_\mathrm{MIT}=2.0428$.  
The free parameters are the bag radius $R$ and the energy splitting 
between the bare nucleon and the bare delta.  
For each $R$, we adjust the splitting such
that the experimental position of the resonance is reproduced.

It is a known drawback of the model that the width of the delta
is underestimated, irrespectively of the bag radius, if the pion
decay constant $f_\pi$ is fixed to the experimental value.
By reducing $f_\pi$ from $93\,\mathrm{MeV}$ to
$83\,\mathrm{MeV} > f_\pi > 78\,\mathrm{MeV}$ 
we are able to reproduce the experimental phase shift in the 
energy range from the threshold to $E\sim 1300\,\mathrm{MeV}$
for $0.8\,\mathrm{fm} < R < 1.1\,\mathrm{fm}$.
Since our aim here is to explore the applicability of the 
method to calculate a wide range of baryon properties as 
measured in pion-production experiments, rather than to 
accurately reproduce particular experimental results, we have 
not attempted to further adjust the parameters of the model.
We keep $R=1.0\,\mathrm{fm}$ and $f_\pi=81\,\mathrm{MeV}$ 
as the standard parameter set.  
Fitting the calculated phase shift with the ansatz (\ref{Kscatter}) 
we get $C =\half\Gamma = 58\,\mathrm{MeV}$ 
and  $D=\tan\delta_\mathrm{b}=-0.42$, where $\delta_\mathrm{b}$
is the background phase shift.
The inclusion of the Roper in (\ref{KN}) contributes less than 5~\% 
to the width.
Taking into account the finite widths of the delta and the Roper
resonances in the evaluation of the sum over intermediate states
(see appendix~A) has a negligible effect on the results.

\begin{figure}[ht]
\begin{center}
\includegraphics[width=84mm]{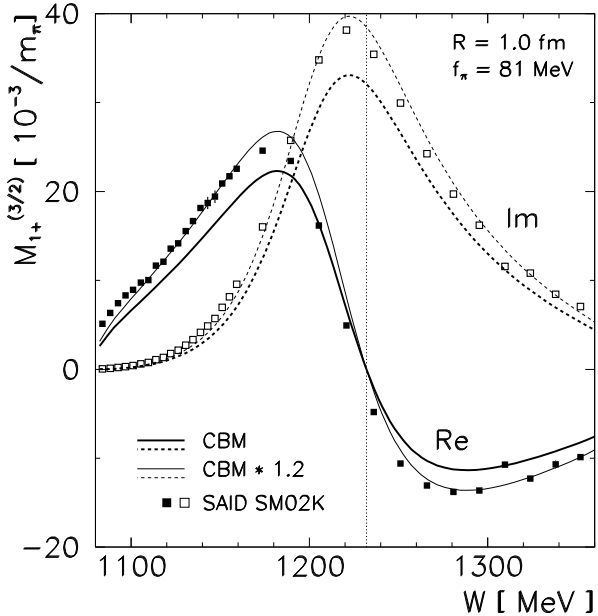}
\end{center}
\caption{The $M_{1+}^{(3/2)}$ electro-production amplitude in the CBM
by using $R=1.0\,\mathrm{fm}$ and $f_\pi=81\,\mathrm{MeV}$ (thick curves),
and multiplied by 1.2 (thin curves).
The data points in the figures are the single-energy values of the SM02K 
($2\,\mathrm{GeV}$) solution of the SAID $\pi\mathrm{N}$ partial-wave
analysis \protect\cite{SAID}.}
\label{fig:M1plus}
\end{figure}

The dominant magnetic contribution calculated from (\ref{M1plus})
is shown in fig.~\ref{fig:M1plus}.
The reason why the experimental values are underestimated 
lies in a too weak $\gamma\mathrm{N}\Delta$ vertex.
In this model it is proportional to the isovector magnetic moment.
For the nucleon its value is typically 20~\% lower than the experimental 
value, almost irrespectively of the model parameters \cite{Theberge}.
Increasing the calculated amplitude by 20~\% we obtain an almost
perfect agreement with the experiment throughout the energy range.

\begin{figure}[ht]
\begin{center}
\includegraphics[width=84mm]{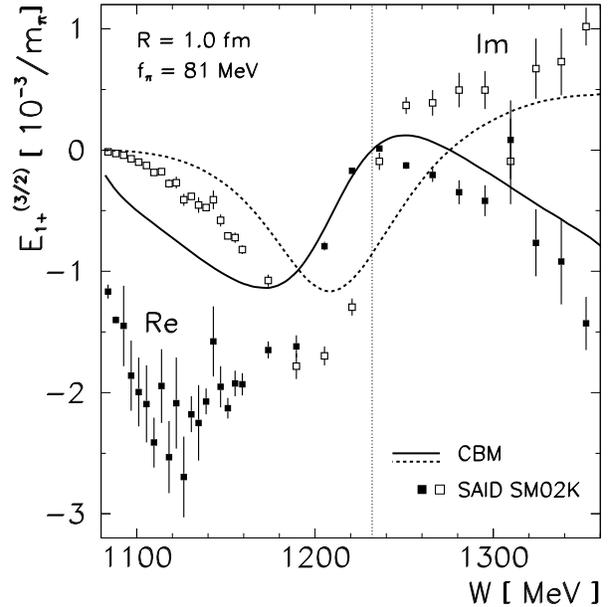}
\end{center}
\caption{The $E_{1+}^{(3/2)}$ electro-production amplitude in the CBM
by using $R=1.0\,\mathrm{fm}$ and $f_\pi=81\,\mathrm{MeV}$.}
\label{fig:E1plus}
\end{figure}

Regarding the $E_{1+}$ amplitude, we encounter the well-known problem 
(see {\it e.g.\/}~\cite{thomas}) of large cancellations of terms 
in the expression for the electro-magnetic current, which leads 
to unreliable results.  Instead, we use current conservation
and calculate $E_{1+}$ from the charge operator.
The energy dependence of the real and imaginary parts
(fig.~\ref{fig:E1plus}) shows the correct pattern compared
to the experiment, though the calculated magnitude is too small.
The agreement is worse at low energies, although the corresponding
experimental uncertainties are large as well.

Since in the $K$-matrix approach we can extract the pure resonance 
contribution at the pole of the $K$ matrix (this would not be possible 
if we worked with the $T$ matrix), we can directly compare our results 
with the calculation of the transition form factors $G_{M1}$ and 
$G_{E2}$ at the photon point within the same model \cite{Tiator88}.
We have explicitly checked that after substituting our matrix elements 
of $V_{mt}$ in eqs.~(\ref{DadadN}) and (\ref{NaD}) by the corresponding 
bare values the results of ref.~\cite{Tiator88} are consistent with ours.
However, while in their calculation of $G_{M1}$ it was possible to 
reproduce the experimental value by reducing the bag radius -- and 
hence increasing the strength of the $\pi$qq vertex -- this mechanism 
does not improve the agreement in the case of the $M_{1+}$ amplitude.
The reason is that increasing the strength of the quark-pion interaction 
leads to a larger width of the resonance, and since $\sqrt{\Gamma}$ 
appears (implicitly) in the denominator of the amplitudes (\ref{Klambda}) 
and (\ref{KL}), $M_{1+}$ {\em decreases\/}.

\begin{figure}[ht]
\begin{center}
\includegraphics[width=84mm]{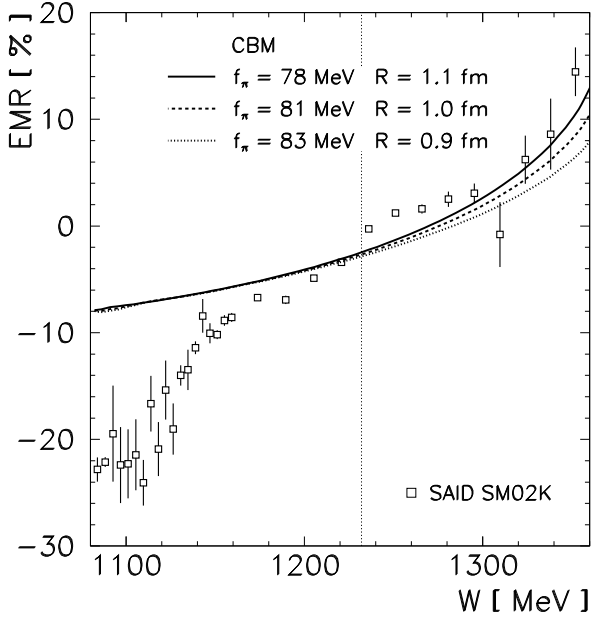}
\end{center}
\caption{The energy dependence of 
$\mathrm{EMR}={\mathrm{Re}\,[\,E_{1+}^{(3/2)\ast} M_{1+}^{(3/2)}\,]/
|\,M_{1+}^{(3/2)}\,|^2}$ at the photon point in the CBM,
for three sets of model parameters.}
\label{fig:EMR}
\end{figure}

In the ratio of the $E_{1+}$ and $M_{1+}$ multipoles (the EMR),
the influence of the too weak $\gamma\mathrm{N}\Delta$ coupling
is strongly reduced, and the agreement with the experiment above
$E\simeq 1150\,\mathrm{MeV}$ is much better (fig.~\ref{fig:EMR}).

\begin{figure}[ht]
\begin{center}
\includegraphics[width=84mm]{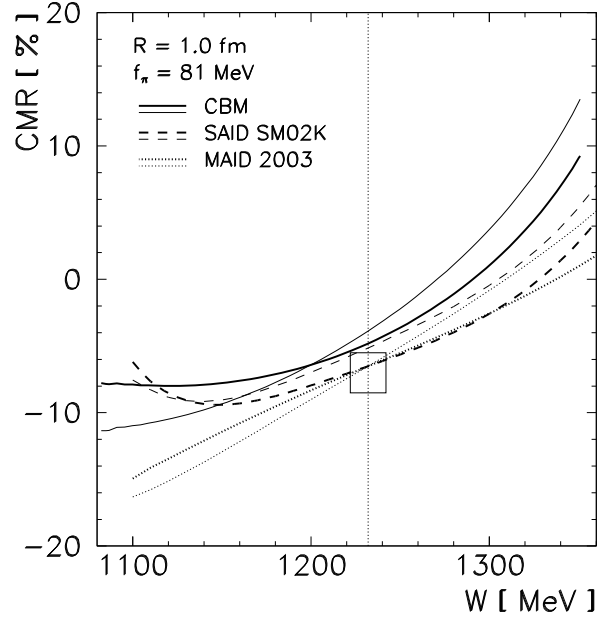}
\end{center}
\caption{The energy dependence of
$\mathrm{CMR}={\mathrm{Re}\,[\,S_{1+}^{(3/2)\ast} M_{1+}^{(3/2)}\,]/
|\,M_{1+}^{(3/2)}\,|^2}$ at $Q^2=0.1$ (thin curves)
and $0.5\,(\mathrm{GeV/c})^2$ (thick curves) in the CBM compared
to the results of SAID and MAID.  The experimental CMR
in the $\Delta E\simeq 10\,\mathrm{MeV}$ vicinity of the $\Delta$
resonance is $\simeq (-7.0\pm 1.5)\,\%$ for $0.1\leq Q^2\leq
0.9\,(\mathrm{GeV/c})^2$ \protect\cite{pospischil,joo} (rectangle).}
\label{fig:CMR}
\end{figure}

In general, the $Q^2$-dependence
of the amplitudes is not well reproduced in the model, partly
due to the rather peculiar form of $v(k)$ at large $k$.
Figure~\ref{fig:CMR} shows the energy dependence of the CMR
for two non-zero values of $Q^2$
compared to SAID \cite{SAID} and MAID \cite{MAID2003} results
based on rather scarce experimental data.  Our calculation reproduces
the general pattern, though the magnitude at the resonance and above it
is not well reproduced.

From our results it is possible to extract the resonance 
parameters at the pole of the $T$~matrix, based on the separation 
of the amplitude into the resonant and background parts, {\it i.e.\/} 
$T=T_\mathrm{R} + T_\mathrm{B}$ using the parameterization
\cite{Tiator96,wilbois} 
$T_\mathrm{R} = r\Gamma_\mathrm{R} \e^{\i\phi}
/(M_\mathrm{R}-E-\i\Gamma_\mathrm{R}/2)$.
The parameters can be expressed in terms of $A$, $B$, $C$, and $D$ 
which are determined by fitting our results to (\ref{Kscatter})
and (\ref{Kepro}).  
Since the parameters of our model were chosen in
order to reproduce the phenomenological phase shift, 
it is not surprising that the pole
of the $T$~matrix appears at 
$E_\mathrm{R} = M_\mathrm{R} - \mathrm{i}\Gamma_\mathrm{R}/2 = 
(1211 - 49\,\mathrm{i})\,\mathrm{MeV}$
which is almost exactly at the correct position 
$(1210 - 50\,\mathrm{i})\,\mathrm{MeV}$ \cite{PDG2004}.
The corresponding moduli and phases for the transverse multipoles
are shown in table~\ref{tab:reiphi}.  While the magnitudes
are underestimated, the ratio as well as the phases
are much better reproduced.

\begin{table*}
\caption{Resonance pole parameters extracted from the computed 
$E_{1+}^{(3/2)}$ and $M_{1+}^{(3/2)}$ multipoles, compared to
various determinations from data.  The moduli $r$ are in units
of $10^{-3}/m_\pi$.}
\begin{center}
\begin{tabular}{lrrrrc}
\hline

\hline

\hline
$R\,[\mathrm{fm}]$/$f_\pi\,[\mathrm{MeV}]$ 
             & $r_E$ & $\phi_E$ & $r_M$ & $\phi_M$ & $R_\Delta$\\
\hline
1.1 / 78 & 0.95 & $-160^\circ$ & 16 &  $-35^\circ$ & $-0.034 - 0.047\,\i$\\
1.0 / 81 & 0.95 & $-165^\circ$ & 16 &  $-38^\circ$ & $-0.035 - 0.047\,\i$\\
0.9 / 83 & 0.97 & $-165^\circ$ & 16 &  $-40^\circ$ & $-0.036 - 0.049\,\i$\\
\hline 
ref.~\protect\cite{Tiator96}
 & $1.23$  & $-154.7^\circ$ & $21.16$ & $-27.5^\circ$ & $-0.035 - 0.046\,\i$ \\
ref.~\protect\cite{wilbois}, MSP fit
 & $1.12$  & $-162^\circ$ & $20.75$ & $-36.5^\circ$ & $-0.040 - 0.047\,\i$ \\
ref.~\protect\cite{Davidson99}, Fit 1 
 & $1.22$ & $-149.7^\circ$  & $22.15$ & $-27.4^\circ$ & $-0.029 - 0.046\,\i$ \\
ref.~\protect\cite{Workman99}, Fit A
 & $1.38$ & $-158^\circ$    & $20.9$  & $-31^\circ$  & $-0.040 - 0.053\,\i$ \\
\hline

\hline

\hline
\end{tabular}
\end{center}
\label{tab:reiphi}
\end{table*}

\section{Summary and conclusions}

We have investigated a method to calculate directly the
$K$~matrices of resonant electro-production processes in 
the framework of chiral quark models.
The main advantage of the method shows up in the treatment of
resonant channels in which the resonant part of the
amplitude can be separated from the background part in an
unambiguous way.  Furthermore, the finite width of
the resonance can be correctly taken into account.

The method has been successfully applied to the calculation
of amplitudes in the $\Delta$ channel in the Cloudy Bag Model.  
In spite of the simplicity of the model we have been able to reproduce
reasonably well the behavior of all amplitudes from the threshold
up to the energies where the two-pion production becomes important.
The method can be applied to other models with more sophisticated 
description of quark dynamics which so far have not been used 
outside the resonance peak. 

In the future we intend to apply the method to the
calculation of electro-production amplitudes in other channels.
Particularly interesting is the Roper channel, where the interplay
between the resonant part induced by the excited quark core
and the background due to the scattering pion being attached
to the nucleon as well as to the delta, becomes crucial.

\vspace{12pt}

This work was supported by the Bilateral Program for Scientific
and Technological Cooperation of the Ministries of Science, Technology,
and Higher Education of Portugal and Slovenia (GRICES and ARRS Agencies).

\appendix

\section{Evaluation of matrix elements}

In models in which the pions are linearly coupled to the quark
source it is possible to derive some general relations for the matrix 
elements, independent of the particular quark model.
Let us first note that if $\Psi_A$ is an eigenstate of the 
Hamiltonian (\ref{Hpi}) then
\begin{equation}
    (\omega_k+H-E_A)a_{mt}(k)|\Psi_A\rangle   =  
  - V^\dagger_{mt}(k)|\Psi_A\rangle\,,
\label{commute1}
\end{equation}
\begin{eqnarray}
   (\omega_k+\omega_k'+H-E_A) a_{mt}(k)a_{m't'}(k')|\Psi_A\rangle =
\nonumber\\ 
&&\kern-180pt  
  - \left[V^\dagger_{mt}(k) a_{m't'}(k') +
    V^\dagger_{m't'}(k')a_{mt}(k)\right]|\Psi_A\rangle\,.
\label{commute2a}
\end{eqnarray}

The renormalization of the operator $\sum_{i=1}^3\sigma_m^i\tau_t^i$
which appears in the quark parts of the EM currents 
(see (\ref{current}) and (\ref{charge})) takes the form
$$
 \langle\Psi^\mathrm{P}_\Delta|| 
          \sum_{i=1}^3{\sigma}^i{\tau}^i||\Phi_\mathrm{N}\rangle
  = {\langle\Psi^\mathrm{P}_\Delta||V(k_0)||\Phi_\mathrm{N}\rangle
          \over v(k_0)}
  = -\sqrt{k_0\over\omega_0}{K_{\pi\pi}(E)\over\pi v(k_0)}
$$
where we have used (\ref{Vmt}), (\ref{KCL}) and  (\ref{K2chi}).

The pion contribution in (\ref{current}) and (\ref{charge}) 
involves two-pion operators.
To illustrate the procedure let us consider the case of two creation
operators: using the conjugate of (\ref{commute2a}) 
and inserting the complete set of states we can write 
\begin{eqnarray}
 \langle\widetilde{\Psi}_\Delta(E)|
         a^\dagger_{mt}(k)a^\dagger_{m't'}(k')|\Phi_\mathrm{N}\rangle
&=&
\nonumber\\
&& \kern-108pt
-  {\langle\Phi_\mathrm{N}| V^\dagger_{m't'}(k')|\Phi_\mathrm{N}\rangle 
  \langle\Phi_\mathrm{N}| a_{mt}(k)|\widetilde{\Psi}_\Delta(E)\rangle
    \over (\omega_k+\omega_k'-\omega_0)}
\nonumber\\
&& \kern-108pt - \sum_{JT}\int{\d E'\,\,K_{\pi\pi}^{JT}(E')^2
                                     \over 1+K_{\pi\pi}^{JT}(E')^2}
   {\langle\Phi_\mathrm{N}| V^\dagger_{m't'}(k')
             |\widetilde{\Psi}_{JT}(E')\rangle 
       \over (\omega_k+\omega_k'-\omega_0)} 
\nonumber\\
&& \kern-108pt
    \times\langle\widetilde{\Psi}_{JT}(E')|
              a_{mt}(k)|\widetilde{\Psi}_\Delta(E)\rangle
   - (k,m,t) \leftrightarrow (k',m',t')\,.
\nonumber\\
\label{DadadN}
\end{eqnarray}
Again, the transition matrix elements involving $a(k)$
and $V(k)$ can be related to the $K$~matrix, {\it e.g.\/}:
\begin{equation}
 \langle\Phi_\mathrm{N}|| a(k)||\Psi^\mathrm{P}_\Delta(E)\rangle
= 
  \delta(k-k_0)
   - { \langle\Phi_\mathrm{N}||V^\dagger(k)||\Psi^\mathrm{P}_{JT}(E)\rangle 
       \over (\omega_k-\omega_0)}\,,
\label{NaD}
\end{equation}
hence
\begin{eqnarray}
 \langle\Phi_\mathrm{N}|| a(k)||\widetilde{\Psi}_\Delta(E)\rangle
&=&
\nonumber \\
 &&\kern-66pt
   K_{\pi\pi}^{-1}\delta(k-k_0)
   - {1\over\pi}\sqrt{k_0\over\omega_0}{\chi(k,k_0)
             \over(\omega_k-\omega_0)\chi(k_0,k_0)}\,.
\nonumber
\end{eqnarray}
The expression $K_{\pi\pi}^{JT}(E')^2/(1+K_{\pi\pi}^{JT}(E')^2)
=\sin^2\delta_{JT}$ is proportional to the cross section in
the $P_{JT}$ channel and can be evaluated either from the calculated
or the experimental phase shift. 
It yields sizable contributions only close to possible resonances 
({\it e.g.\/} the delta and the Roper).
Furthermore, for a sufficiently narrow resonance at\break
$E'=E_*$, this expression can be substituted by\break 
${1\over2}\pi\Gamma\delta(E'-E_*)$
leading to a similar expression as in the perturbation theory.
As a consequence, the matrix element $\langle\widetilde{\Psi}_{JT}(E')|
a_{mt}(k)|\widetilde{\Psi}_\Delta(E)\rangle$
in the last term substantially contributes  only for $JT=\thalf\thalf$ 
and $E'\approx E_\Delta$.

A similar procedure is used to extract the one-pion amplitude
around the bare delta below the 2-pion threshold:
\begin{eqnarray}
\langle\Delta ||a(k)||\Psi_\Delta(E)\rangle
&=& -\int_{E_N+m_\pi}^\infty \d E'\, \sqrt{\omega_0'\over k_0'}\,
\,{K_{\pi\pi}^{JT}(E')^2 \over 1+K_{\pi\pi}^{JT}(E')^2}
\nonumber \\
  &&\times {\langle\widetilde{\Psi}_\Delta(E') ||V^\dagger(k)
        ||\widetilde{\Psi}_\Delta(E)\rangle \over (\omega_k+E'-E)}\,.
\nonumber
\end{eqnarray}

\end{document}